\documentclass[reprint, amsmath,amssymb, aps,superscriptaddress,floatfix,prx]{revtex4-2} 
\pdfoutput=1
\usepackage{graphicx}

\usepackage[utf8]{inputenc}
\usepackage{xcolor}
\usepackage{chemformula}
\definecolor{codegreen}{rgb}{0,0.6,0}
\definecolor{codegray}{rgb}{0.5,0.5,0.5}
\definecolor{codepurple}{rgb}{0.58,0,0.82}
\definecolor{backcolour}{rgb}{0.95,0.95,0.92}
\definecolor{urlblue}{HTML}{007bff}

\usepackage{dcolumn}
\usepackage{bm}
\usepackage{bbm}
\usepackage{orcidlink}
\usepackage{cleveref}
\usepackage{microtype}
\usepackage{braket}
\usepackage{mathtools}

\newcommand{\Tr}{\ensuremath{\textup{Tr}}}

\newcommand{\1}{\ensuremath{\mathbbm{1}}}

\hypersetup{
    colorlinks = true,
    linkcolor  = urlblue,
    citecolor  = urlblue,
    urlcolor   = urlblue,
}

\usepackage[mathlines]{lineno}

\begin{document}

\title{Kitaev-Heisenberg model on the star lattice -- from chiral Majoranas to chiral triplons}

\author{P. d'Ornellas\orcidlink{0000-0002-2349-0044}}
\affiliation{\small Blackett Laboratory, Imperial College London, London SW7 2AZ, United Kingdom}

\author{J. Knolle\orcidlink{0000-0002-0956-2419}}
\affiliation{\small Blackett Laboratory, Imperial College London, London SW7 2AZ, United Kingdom}
\affiliation{Department of Physics TQM, Technische Universit{\"a}t M{\"u}nchen, James-Franck-Stra{\ss}e 1, D-85748 Garching, Germany}
\affiliation{Munich Center for Quantum Science and Technology (MCQST), 80799 Munich, Germany}

\date{\today}

\begin{abstract}

The interplay of frustrated interactions and lattice geometry can lead to a variety of exotic quantum phases. Here we unearth a particularly rich phase diagram of the Kitaev-Heisenberg model on the star lattice, a triangle decorated honeycomb lattice breaking sublattice symmetry. In the antiferromagnetic regime, the interplay of Heisenberg coupling and geometric frustration leads to the formation of valence bond solid (VBS) phases -- a singlet VBS and a bond selective triplet VBS stabilized by the Kitaev exchange. We show that the ratio of the Kitaev versus Heisenberg exchange tunes between these VBS phases and chiral quantum spin liquid regimes. Remarkably, the VBS phases host a whole variety of chiral triplon excitations with high Chern numbers in the presence of a weak magnetic field. We discuss our results in light of a recently synthesized star lattice material and other decorated lattice systems. 
\end{abstract}

\maketitle


\section{Introduction}

A central theme in the study of quantum magnetism has been the discovery and classification of materials hosting different types of collective excitations. This has unearthed a rich variety of ground state phases which can serve as non-trivial vacua, for example, for chiral topological excitations. Such excitations are promising for spintronics applications~\cite{wang2018topological} and are realized in a variety of magnetic phases: First,  topological magnon insulators (TMIs) are conventional long-range spin ordered magnets whose bulk magnon bands have nonzero topological invariants. As a result they can display magnon surface states with uni-directional propagation~\cite{Katsura_2010,zhang2013topological,Mook_2014,mcclarty2022topological}. 
A second example beyond conventional ordered magnets are valence bond solids (VBS) \cite{auerbach_interacting_1994}, with short range spin correlations. Rather, neighbouring spins form highly entangled pairs and the corresponding dimer correlation function is long-range ordered. In particular, in a singlet VBS state neighbouring spins approximate a spin singlet, forming a gapped dimer ground state that may host chiral boundary modes of triplon excitations.
These were first predicted for the Shastry–Sutherland model system SrCu$_2$(BO$_3$)$_2$, where the addition of Dzyaloshinskii–Moriya (DM) interactions leads to a spin-orbit type coupling of triplons, such that gaps induced by a small magnetic field endow the triplon bands with a non-zero Chern number~\cite{Romhanyi_2015, Mcclarty_2017,malki2017magnetic}.
The third and most exotic example are quantum spin liquids (QSLs) with topologically ordered ground states~\cite{savary2016quantum,zhou2017quantum,knolle2019field}. Chiral QSLs with broken time reversal symmetry (TRS) can host fractionalized surface modes with unidirectional propagation akin to their electronic brethren of the (fractional) quantum Hall effect~\cite{kalmeyer1987equivalence}. 

\begin{figure*}[t]
    \centering
    \includegraphics[width=0.9\textwidth]{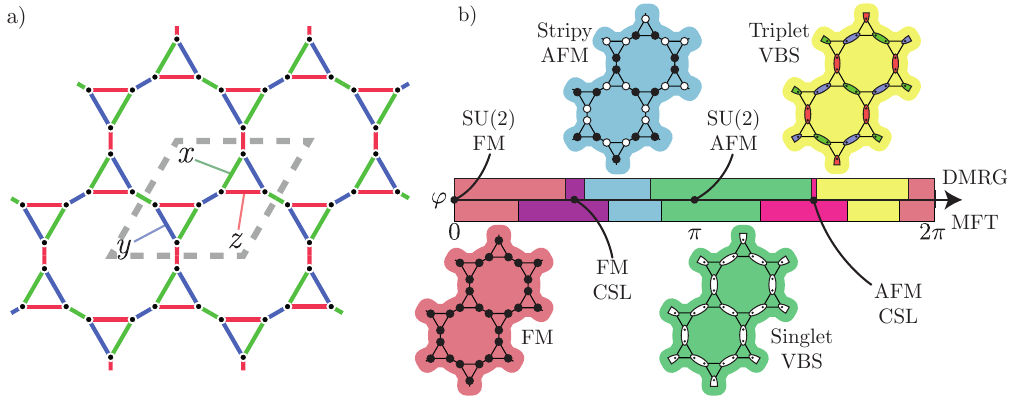}
    \caption{ 
    \textbf{a)} A fragment of the star lattice is shown, with three colours to indicate the assignment of bond labels $x$, $y$ and $z$ for the Kitaev interaction given in the Hamiltonian, Eqn.~\ref{eqn:hamiltonian}. \textbf{b)} The derived phase diagram as a function of the parameter $\varphi$ defined in Eqn.~\ref{eqn:phi_def}. Six phases are shown. At $\varphi = \pi/2$ and $\varphi = 3\pi/2$ the system forms a ferromagnetic CSL and an antiferromagnetic CSL respectively. Two magnetically ordered phases are found around $\varphi < \pi$, where the system forms ferromagnetic or stripy order, with the SU(2)-symmetric, pure Heisenberg model appearing at $\varphi = 0$. Finally, two VBS are formed -- a spin singlet VBS around $\varphi \sim \pi$, where the exact SU(2)-symmetric antiferromagnetic Heisenberg model is found at $\varphi = \pi$, and an unusual triplet VBS phase around $\varphi = 1.7\pi$, in which the system condenses into a set of spin triplets, where the combined spin-1 state is given by $\ket {t_\alpha}$ defined in Eqn.~\ref{eqn:tx_def}-\ref{eqn:tz_def}.
    }
    \label{fig:lattice_and_phase_diagram}
\end{figure*}

Finding stable chiral excitations in microscopic models of insulating magnets has turned out to be a major challenge. Often, the ground states themselves are very fragile, as has been the case for many QSLs~\cite{szasz2020chiral,chen2022quantum}, or the topological edge excitations may be unstable to decay processes from interactions~\cite{chernyshev2016damped,habel2023breakdown}. In this context, the Kitaev honeycomb model with its bond-anisotropic Ising interactions stands out as a rigorous example of a stable QSL~\cite{Kitaev_2006} and a robust TMI phase in large magnetic fields~\cite{mcclarty2018topological,joshi2018topological}. For small magnetic field the gapless Kitaev QSL turns into a gapped QSL with chiral Majorana boundary modes. Moreover, the presence of perturbing interactions like Heisenberg exchange present in any real materials' realization gives rise to a rich phase diagram with various long-range ordered magnetic phases~\cite{Chaloupka_2013,rau2016spin} and, surprisingly, TMI behavior in the field-polarized regime~\cite{mcclarty2018topological,joshi2018topological}. Hence, excitations of the (extended) Kitaev honeycomb model can be tuned from chiral Majorana excitations to stable chiral magnons as a function of increasing magnetic field. This is not only interesting theoretically but of direct experimental relevance because chiral magnetic excitations give rise to a thermal Hall response~\cite{Katsura_2010,Matsumoto_2011}. The latter has been measured in the Kitaev magnet $\alpha$-RuCl$_3$~\cite{kasahara2018majorana,bruin2022robustness}, but whether its origin is related to QSL or TMI behavior is still under debate~\cite{czajka2023planar,chern2021sign}.  

Following Kitaev's original proposal for the honeycomb lattice~\cite{Kitaev_2006} subsequent work established that his model can be exactly solved on a wide variety of lattices with coordination number three, e.g.~different 2D and 3D lattices ~\cite{Yao_2007,mandal2009exactly,hermanns2014quantum,o2016classification,peri_non-abelian_2020} and even disordered amorphous lattices~\cite{Cassella_2023,grushin2023amorphous}. 
A particularly interesting case is the star lattice -- a honeycomb lattice with each vertex decorated by a triangle as shown in Fig.~\ref{fig:lattice_and_phase_diagram} -- which is one of the eleven Archimedian tilings in 2D~\cite{richter2004quantum}. There, the Kitaev QSL was the first rigorous example of a chiral QSL breaking TRS {\it spontaneously}~\cite{Yao_2007,dusuel_perturbative_2008}. The presence of plaquettes with an odd number of sites leads to broken TRS in a given flux configuration, such that the zero flux ground state is a chiral QSL with Majorana edge modes. Despite much research on the stability of the Kitaev QSL on the honeycomb lattice~\cite{winter2017models,hermanns2018physics,Trebst_2022} the effect of perturbing interactions on the chiral Kitaev QSL has not been investigated on the star lattice. This is even more surprising given that the pure antiferromagnetic Heisenberg model on the star lattice is one of the rare established examples hosting a singlet VBS ground state phase with triplon excitations~\cite{Richter_2004,choy2009classification,Yang_2010, Jahromi_2018,jahromi2018infinite,ran2018emergent,reingruber2023thermodynamics}. Additional motivation comes from the recent synthesis of the first $S=1/2$ materials realisation of the star lattice~\cite{sorolla_synthesis_2020,ji2022dielectric} (previous star lattice realisations contained $S=5/2$ spin clusters~\cite{zheng2007star}).

In this paper we investigate the rich ground state phase diagram and distinct types of chiral excitations of the  Kitaev-Heisenberg model on the star lattice. We show that, in addition to the chiral Kitaev QSL and various magnetically ordered phases, a novel bond-anisotropic triplet VBS is realized. The triplon bands show non-zero Chern numbers once a small magnetic field is included to induce gaps. Thus, we establish that -- in analogy to the honeycomb Kitaev model, which can be tuned with a magnetic field from hosting chiral Majorana to chiral magnon excitations -- the Kitaev-Heisenberg model on the star lattice tunes from chiral Majoranas to chiral triplons. 

One of the competing phases that can form in lieu of magnetic order or QSLs are VBSs. In their most common form, neighbouring pairs of spins anti-align into singlets, forming a dimer covering of the entire lattice. On this background, excitations involve breaking a singlet state and creating a spin-1 triplet -- a so-called triplon mode that can be represented as a bosonic particle~\cite{Sachdev_1990}. Recently, interest has been generated in triplons as a way of realising a bosonic bulk-boundary correspondence. This has been proposed both in 1D topological chain systems~\cite{joshi_topological_2017}, as well as 2D bosonic analogues of the quantum Hall effect~\cite{Romhanyi_2011,Romhanyi_2015,Mcclarty_2017, Nawa_2019, malki_magnetic_2017}. In general, such effects have relied on the presence of Dzyaloshinskii–Moriya (DM) interactions to generate complex hoppings in the triplon Hamiltonian, allowing for non-zero Chern and winding numbers to appear. Here, we will show that the Kitaev interaction provides an alternative route for a triplon Hall state forming in the absence of DM interactions. For the Heisenberg-Kitaev model on the star lattice we unearth an extremely rich phase diagram containing singlet and triplet VBSs with a wide range of Chern numbers.

The structure of the paper is as follows. 
In \textsection\ref{sec:the_model} we introduce the model, describing the parameters used to tune the system and discussing previous studies that have looked at both the star and the honeycomb lattices. Following this, our investigation is composed of three parts. 
In \textsection\ref{sec:dmrg} we start by numerically characterising the ground state phase diagram using infinite density matrix renormalisation group (iDMRG) methods. In particular we find two phases in the star lattice that do not have an analogue on the Honeycomb lattice, where geometric frustration leads to VBS ground states. We discuss the chiral Majorana excitations of the Kitaev QSL phase within a parton mean-field theory (MFT).
In \textsection\ref{sec:spin_flip_transformation} we describe a \textit{four-unit-cell spin rotation} that reveals a hidden isometry of the ground state phase diagram. We show that the four non-spin liquid phases may be categorised into two sets, which are equivalent under this transformation.
Finally, in \textsection\ref{sec:bond_operator} we  construct a bosonic mean field theory describing the triplon excitations of the VBS phase. The topological properties of the triplon phases are derived and we confirm the existence of topologically protected triplon edge modes. We conclude in \textsection\ref{Discussion:Outlook} with a discussion of experimental relevance and open questions for future research.

\section{The Kitaev-Heisenberg Model} 
\label{sec:the_model}

Since Kitaev's original proposal, immense effort has been expended to find a physical candidate capable of hosting such a QSL phase, leading to the discovery of a number of so called \textit{Kitaev materials}~\cite{winter2017models,hermanns2018physics,Trebst_2022}, such as \ch{Na2IrO3} \cite{Gretarsson_2013, Hwan_2015} and $\alpha$-\ch{RuCl3} \cite{Sandilands_2015,Nasu_2016,Banerjee_proximate_2016}. Although the exact microscopics are still under debate, there is evidence that these materials realise dominant Kitaev-type interactions, along with subdominant Heisenberg coupling terms which inhibit the formation of a spin liquid ground state. Thus, the Kitaev-Heisenberg (KH) model has been proposed and extensively studied as a potential description of these materials \cite{Chaloupka_2010,Chaloupka_2013,Jiang_2011, Reuther_2011, Schaffer_2012,Price_2012}. On the honeycomb, the full KH phase diagram contains two spin liquid phases and four competing magnetically ordered phases. Despite the insensitivity of the spin liquid to lattice geometry, two of the competing phases found on the honeycomb cannot be constructed on a lattice without sublattice symmetry. 
These phases, namely the N\'eel and stripe order, fall into a broadly antiferromagnetic region of the phase diagram, and thus are geometrically frustrated when the lattice contains odd cycles.

\begin{figure}[t]
    \centering
    \includegraphics[width=\columnwidth]{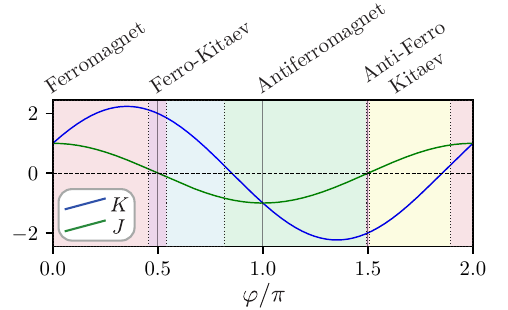}
    \caption{
        The dependence of the $J$ and $K$ values on the parameter $\varphi$ is shown. The two pure Heisenberg and two pure Kitaev points are labelled. DMRG phase boundaries are shown in the background colour.
    }
    \label{fig:jk_diagram}
\end{figure}

We start by defining the model on the star lattice, shown in Fig.~\ref{fig:lattice_and_phase_diagram}a. On each site is a single spin-1/2, and bonds are `coloured' with a label $\alpha \in \{x,y,z\}$ such that no vertex touches two bonds of the same colour.
We choose the Hamiltonian to interpolate between two well-studied limits: the spin isotropic Heisenberg and the anisotropic Kitaev limit where each bond of type $\alpha$ is coupled only  in the direction that matches the colouring: 
\begin{align} \label{eqn:hamiltonian}
    H = -
    \sum_{\braket{jk}_\alpha } \bigg (  K \sigma_j^\alpha \sigma_k^\alpha
    +J \sum_{\bar \alpha \neq \alpha}  \sigma_j^{\bar \alpha} \sigma_k^{\bar \alpha}
    \bigg ).
\end{align}
Here $\braket{jk}_\alpha$ indicates a bond of type $\alpha$ that connects adjacent sites $j$ and $k$. The parameter $K$ determines the strength of the ``on-colouring'' coupling, where the coupling direction matches $\alpha$, and $J$ determines the strength of the two remaining ``off-colouring'' couplings. In analogy with previous studies~\cite{Chaloupka_2013, Gohlke_2017}, let us introduce a single parameter $\varphi \in [0,2\pi]$ that smoothly maps onto the possibilities for $K$ and $J$ according to
\begin{align} \begin{aligned} \label{eqn:phi_def}
    K &= 2\sin \varphi + \cos \varphi, \\
    J &= \cos \varphi,
\end{aligned}\end{align}
shown in Fig.~\ref{fig:jk_diagram}. Here, two points where $J = K$ can be found at $\varphi = 0$ and $\pi$, representing the two SU(2)-symmetric points where the system exactly describes a ferromagnetic or antiferromagnetic Heisenberg model.

\section{Ground State Phase Diagram} \label{sec:dmrg}
We briefly recap the phase diagram of the standard Kitaev model on the honeycomb lattice~\cite{Chaloupka_2013, Gohlke_2017}, which will be useful for elucidating the novel features of the star lattice. The full phase diagram in $\varphi$ contains the two Kitaev spin liquid phases, and four distinct magnetically ordered phases. These come in two pairs, where the two phases in each pair can be mapped exactly onto one another by rotating the coordinate system of certain spins in the lattice~\cite{Schaffer_2012, Chaloupka_2010,kimchi2014kitaev} according to a four unit cell pattern. 

Around $\varphi = 0$ ($K, J >0$) the system orders into a ferromagnetic (FM) phase, with completely aligned spins. Around $\varphi = 0.6\pi$ ($K>0, J<0$) the system forms a stripy ordered phase, which can be mapped onto the ferromagnetic phase by the four unit cell spin rotation. Around $\varphi = \pi$ ($K,J < 0$) the system forms an antiferromagnetic (AFM) N\'eel ordered phase, which is again isomorphic to a zigzag phase that appears around $\varphi = 1.6\pi$ ($K<0, J>0$). The mapping implies that the dynamics in each pair of isomorphic phases is identical.



Next, let us turn our attention to the star lattice, We numerically determine the ground state across the full range of $\varphi$ using the density matrix renormalisation group (DMRG) method. DMRG is a variational optimisation algorithm used to find the ground state of many-body Hamiltonians. Although it is predominantly used for characterising 1D physics, recent work has shown that it can provide an effective description of several 2D frustrated systems~\cite{Gohlke_2017,Shinjo_2015}. Here, we construct the Hamiltonian on a system with infinite cylindrical geometry, with two unit cells (12 sites) spanning the finite circumference. The ground state is represented with a matrix product state ansatz with bond dimension $\chi = 1600$, and is calculated using the TeNPy library~\cite{tenpy} for 1000 values of $\varphi$ across the full range $[0,2\pi]$.


Using this method, we find that the phase diagram contains six distinct phases, shown in Fig.~\ref{fig:lattice_and_phase_diagram}b. In the regime where $K>0$ ($\varphi \sim [-0.2\pi, 0.8\pi]$) -- which we will refer to as the broadly FM regime -- we find two magnetically ordered phases. Around $\varphi = 0$, we encounter the standard FM phase with aligned spins, which includes the FM Heisenberg point at $\varphi = 0$. 

Around $\varphi = 0.6$ we find a stripy phase with an equal proportion of spin up and spin down sites. This phase may be understood in direct analogy with the stripy ordered phase found in the honeycomb model, where each triangle in the star lattice can be treated as an effective spin half site, since they form either a $\ket{\uparrow_{\textup{eff}}} = \ket{\uparrow \uparrow \downarrow}$ or a $\ket{\downarrow_{\textup{eff}}} = \ket{\uparrow \downarrow \downarrow}$ state. These effective spin half triangles then order identically to the stripe order found on the honeycomb. Remarkably, we also find that the FM and stripy ground states can be mapped onto  one under an extended four unit cell rotation described in \textsection\ref{sec:spin_flip_transformation}.

In the regime where $K<0$ ($\varphi \sim [0.8\pi, 1.8\pi]$) -- referred to as the broadly AFM regime -- the system forms two distinct VBS phases. Around $\varphi = \pi$ we have a singlet VBS, where the states form spin singlets
\begin{align}
    \ket{s} = \frac{1}{\sqrt 2} \left ( \ket{\uparrow \downarrow}-\ket{ \downarrow \uparrow} \right ), \label{eqn:s_def}
\end{align}
on the inter-triangle bonds. At the point $\varphi = \pi$ the system is exactly an antiferromagnetic Heisenberg model, where previous studies support our conclusions \cite{Richter_2004, Yang_2010, Jahromi_2018}. This VBS phase has no analogue on the honeycomb lattice, which forms a N\'eel ordered ground state. This is due to geometric frustration, where the presence of odd cycles on the star lattice disallows the formation of the N\'eel AFM phase. 

Around $\varphi = 1.7\pi$, the system forms an exotic spin triplet VBS. Here the spins on each bond pair into one of three triplet states depending on the bond type. Bonds of type $x$, $y$ or $z$ form a $\ket{t_x}$, $\ket{t_y}$ or $\ket{t_z}$ state respectively, defined as
\begin{align}
    \ket{t_x} = \frac{i}{\sqrt 2} \left ( \ket{\uparrow \uparrow}-\ket{\downarrow \downarrow} \right ), \label{eqn:tx_def} \\
    \ket{t_y} = \frac{1}{\sqrt 2} \left ( \ket{\uparrow \uparrow}+\ket{\downarrow \downarrow} \right ),\label{eqn:ty_def} \\
    \ket{t_z} =  \frac{-i}{\sqrt 2} \left ( \ket{\uparrow \downarrow}+\ket{\downarrow \uparrow} \right ).\label{eqn:tz_def}
\end{align}
Note that the factors of $i$ have been chosen such that the singlet and triplet states have time reversal symmetry.
We will show below in in \textsection\ref{sec:spin_flip_transformation} that the singlet and triplet VBS phases are isomorphic under a generalized four unit cell spin rotation.

Around $\varphi \sim 0.5\pi$ and $\varphi \sim 1.5\pi$ the system forms two extended chiral spin liquid phases. We notes that the AFM Kitaev phase is stable over a much smaller region of phase space than the FM QSL. This can be explained by noticing that the two competing phases (both the VBS phases) have lower energy than their ferromagnetic counterparts, whereas the energy scale of both Kitaev phases are identical. Thus, as we tune across the phase diagram, the VBS phases very quickly become energetically favourable over the QSL.

As in the honeycomb case, all phase transitions corresponded to cusps in the ground state energy, indicating that the transitions are first order. However, it remains an open and numerically challenging question whether they remain first order in the thermodynamic limit. Full iDMRG results are discussed in Appendix~\ref{apx:DMRG_results}.

\begin{figure}[t]
    \centering
    \includegraphics[width=0.9\columnwidth]{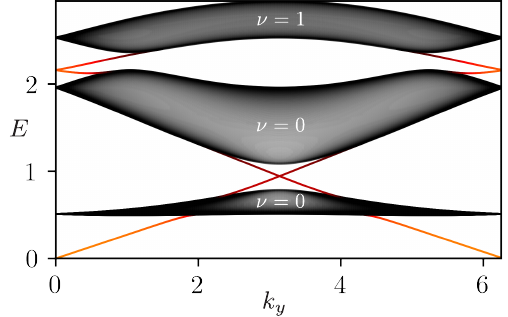}
    \caption{The Majorana fermion band structure calculated for the chiral QSL phase within a parton MFT for a strip-geometry, with periodic boundaries in the $x$-direction and open boundaries in the $y$ direction. We have tuned the model to the antiferromagnetic Kitaev regime ($\alpha = 1.45\pi$) in the vicinity of the soluble point. Energies are plotted as a function of momentum in the $y$-direction, for a sample with $L_x = L_y = 200$. The Chern number $\nu$ of the three bands is labelled. States are coloured according to their inverse participation ratio, where red indicates a localised edge mode and grey a delocalised bulk mode. Note that we only plot the positive-energy Majorana modes, however a set of three `occupied' negative-energy bands are also present, related to the positive-energy bands by particle hole symmetry. Each of these reflected bands has the opposite Chern number to its positive-energy counterpart.}
    \label{fig:kitaev_chern_bands}
\end{figure}
Finally, we have confirmed that the pure Kitaev points ($\varphi = 0.5\pi$ or $1.5\pi$) exactly recover the chiral QSL studied Yao and Kivelson~\cite{Yao_2007}. Here, the model is exactly solvable and can be decoupled into non-interacting Majorana fermions in the presence of a static $\mathbb Z_2$ gauge field~\cite{Kitaev_2006}. However, as soon as we tune away from these points (and some Heisenberg interactions are present) the $\mathbb Z_2$ gauge field is no longer conserved and the system ceases to be exactly solvable. In Appendix \ref{sec:mft}, we study the entire phase diagram using a Majorana MFT~\cite{Schaffer_2012,Knolle_2018}. As shown in Fig.~\ref{fig:lattice_and_phase_diagram}b we recover the DMRG phase diagram qualitatively but, as common for parton MFT, we find an enhancement of the region of stability of the QSL regimes.  
Using this Majorana MFT we are able to obtain the excitation spectrum of the chiral QSL away from the two points where the model is exactly soluble. In Fig.~\ref{fig:kitaev_chern_bands} we show the Majorana band structure on a cylinder geometry with open boundary conditions. The nonzero Chern numbers of the bulk bands give rise to chiral edge modes (indicated in red), whose dispersions are weakly renormalized by the presence of the Heisenberg interactions. 

\section{Isometries of the Kitaev-Heisenberg Phase Diagram}\label{sec:spin_flip_transformation}
\begin{figure}[t]
    \centering
    \includegraphics[width=\columnwidth]{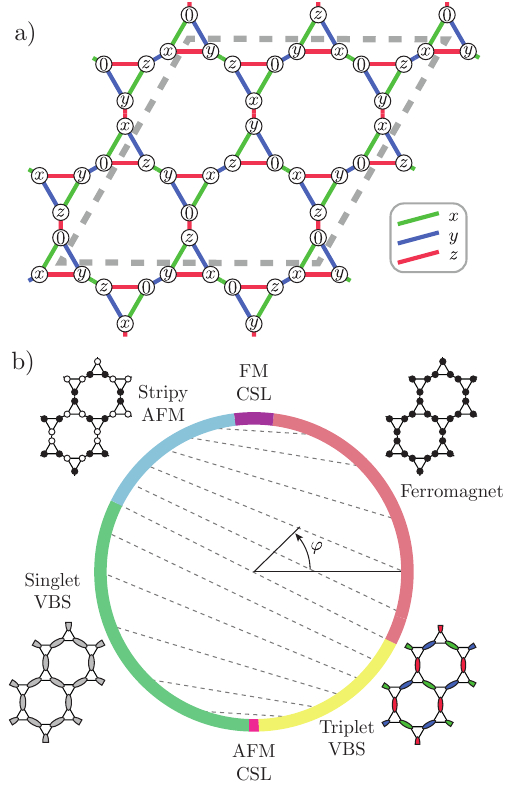}
    \caption{
    \textbf{a)} Plot of the four unit cell transformation that encodes the isometry of the KH phase diagram. Each $0$-marked site is untouched, however the sites labelled with $x$, $y$ or $z$ each experience a $\pi$-rotation around the labelled axis in spin space. Bond-types are indicated by colour. Enlarged unit cell is depicted with the dashed rhombus. \textbf{b)} The KH phase diagram as parametrised by $\varphi$ defined in Eqn.~\ref{eqn:phi_def}. Dashed lines indicate points that are isomorphic under the spin rotation. 
    }
    \label{fig:Appendix_transformation}
\end{figure}

One of the remarkable aspects of the Heisenberg-Kitaev model on the honeycomb lattice is a hidden symmetry between different phases of the model~\cite{Chaloupka_2013,kimchi2014kitaev}. In this section we show that similar relations hold on the star lattice, which help to understand the phase diagram and clarify the connection between the distinct types of VBS phases. 

Let us introduce a spin transformation by dividing the spins in the system into four sets, each labelled with $0$, $x$, $y$ or $z$, arranged in a four-unit-cell pattern as shown in Fig.~\ref{fig:Appendix_transformation}a. Spins marked with $0$ will be left untouched, however on the other spins we apply a $\pi$-rotation in spin space, where the axis of rotation is determined by the site labels. Under such a rotation, the sign of the two spin operators that do not line up with the rotational axis will be flipped, for example if the site $j$ is labelled with $x$ we have
\begin{align}
    \begin{aligned}
        \sigma_j^x &\rightarrow \sigma_j^x \\
        \sigma_j^y &\rightarrow -\sigma_j^y \\
        \sigma_j^z &\rightarrow -\sigma_j^z,
    \end{aligned}
\end{align}
with a similar relation for $y$ and $z$-labelled sites.

Under this transformation, we note that every $x$-bond is always bordered by either a $0 - x$ pair of sites or a $y-z$ pair. Thus, every $x$-type bond operator in $H$ transforms according to
\begin{align}
    \begin{aligned}
        \sigma_j^x\sigma_k^x &\rightarrow \sigma_j^x\sigma_k^x \\
        \sigma_j^y\sigma_k^y &\rightarrow -\sigma_j^y\sigma_k^y \\
        \sigma_j^z\sigma_k^y &\rightarrow -\sigma_j^z\sigma_k^y.
    \end{aligned}
\end{align}
Similarly, $y$-bonds are bordered by $0-y$ or $x-z$, and $z$-bonds are bordered by $0-z$ or $x-y$. These both transform analogously under the spin rotation, where the on-colouring direction remains unchanged and the off-colouring directions have their sign flipped. Thus, our spin rotation is equivalent to transforming the parameters of the Hamiltonian according to
\begin{align}
    K \rightarrow K,\ J \rightarrow -J.
\end{align}
In terms of the single parameter $\varphi$, this is equivalent to transforming to a new $\tilde \varphi$ that satisfies
\begin{align}
     \tan \tilde \varphi = -\tan \varphi - 1
\end{align}
with the correspondence indicated with dashed lines in Fig.~\ref{fig:Appendix_transformation}b.

Thus, we see that there is a correspondence between the ferromagnetic Hamiltonian and the stripy ordered Hamiltonian, as well as a correspondence between the singlet VBS and the stripy VBS Hamiltonians. We can similarly verify that the transformation gives the correct relation between the phases: 
First, The FM case is easy -- starting with a FM alignment in the $z$-direction, we can see that any site labelled with $x$ or $y$ has its spin flipped, which leads precisely to the stripy magnetically ordered phase. Second, we look at the more subtle VBS correspondence. A spin singlet on an $x$-labelled bond will be adjacent to either a $0-x$ pair of sites, in which case the resulting state will be transformed as
\begin{align}
    \ket s  \sim 
    \ket{\uparrow \downarrow} - \ket{\downarrow \uparrow} 
    \rightarrow 
    \ket{\uparrow \uparrow} - \ket{\downarrow \downarrow} 
    \sim \ket{t_x},
\end{align}
or the bond will border two sites of type $y-z$, in which case the transformation is
\begin{align}
    \ket s  \sim 
    \ket{\uparrow \downarrow} - \ket{\downarrow \uparrow} 
    \rightarrow 
    i \ket{\downarrow \downarrow} - i \ket{\uparrow \uparrow} 
    \sim \ket{t_x}.
\end{align}
In both cases we see that (up to a phase) the singlet is mapped onto a triplet $\ket{t_x}$ state. Similar arguments apply for $y$- and $z$-bonds. Thus, from the mapping we observe that the resulting stripy VBS consists of a condensate of triplet states, where each $x$-bond has a $\ket{t_x}$ triplet, each $y$-bond has a $\ket{t_y}$ triplet and so on. This is perfectly consistent with our results in section \ref{sec:dmrg} and the Appendix \ref{sec:mft}.

\section{Triplon Excitations in the VBS}\label{sec:bond_operator}

\begin{figure}[t]
    \centering
    \includegraphics[width=0.8\columnwidth]{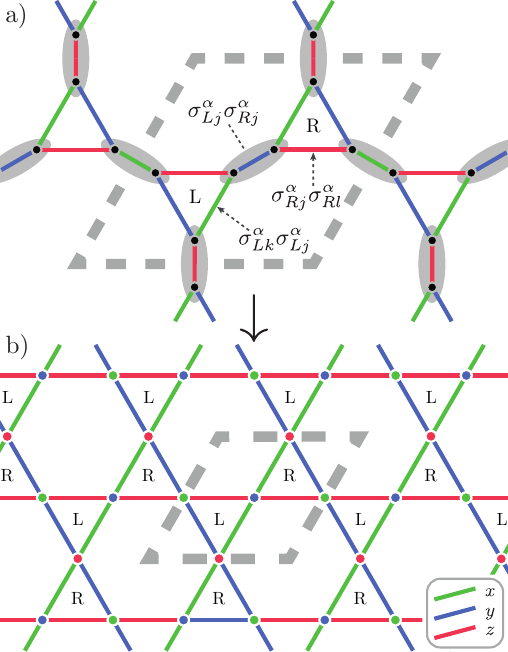}
    \caption{ 
    \textbf{a)} Diagram of the lattice with positions of spin singlets highlighted in grey. Spins are labelled as $\sigma_{R/L}$ depending on which triangle they form a part of in the unit cell. \textbf{b)} The derived Kagome lattice once all the singlets are collapsed to a single site. Here, both edges and sites have a Kitaev labelling $\alpha \in {x,y,z}$, denoted by colour. }
    \label{fig:lattice_to_kagome}
\end{figure}

Having found an exact mapping between the VBS phases we now focus on the singlet VBS  around $\varphi = \pi$. Our objective is to find an effective description the phase and analyse its excitation spectrum. The numerical calculation indicate that spin singlets form on the inter-triangle bonds, thus we shall start by pairing the spins on these bonds and transform to the bond operator representation introduced by Sachdev and Bhatt~\cite{Sachdev_1990, Gopalan_1994}. Spins on a given bond are labelled with $\sigma_L$ and $\sigma_R$, and treated as a single site, collapsing the star lattice into a Kagome lattice, as shown in Fig.~\ref{fig:lattice_to_kagome}. We can then rewrite the Hamiltonian in a two-spin basis for each dimerised bond, given by Eqns.~\ref{eqn:s_def}-\ref{eqn:tz_def}.

In order to apply a MFT, let us re-express these states in terms of bosonic creation operators. Starting with a vacuum $\ket 0$ which contains no particles whatsoever, we define four bosonic creation operators that create singlet and triplet states,
$ s^\dag \ket 0 = \ket s$, 
$ t_\alpha^\dag \ket 0 = \ket {t_\alpha}$, which satisfy the relevant commutation relations,
$[s, s^\dag] = 1$, 
$[t_\alpha, t_\beta^\dag] = \delta_{\alpha \beta}$, 
$[s, t_\alpha^\dag] = 0$. In this basis, the spin operators take the following form
\begin{align}
    \sigma_L^\alpha &= -i(s^\dag t_\alpha - t^\dag_\alpha s) - i\varepsilon _{\alpha \beta \gamma} t^\dag_\beta t_\gamma,\\ 
    \sigma_R^\alpha &= i(s^\dag t_\alpha - t^\dag_\alpha s) - i\varepsilon _{\alpha \beta \gamma} t^\dag_\beta t_\gamma.
\end{align}
Of course, in doing so we have artificially enlarged our Hilbert space. For a state to be legitimate, all sites must be populated by a single boson -- any state with empty or multiply occupied sites will be in the `unphysical' part of the extended Hilbert space. Thus, we introduce a chemical potential term to $H$ that enforces single occupancy per site on average,
\begin{align}
    H_\mu = - \sum_{j \in  \mathcal D}
    \mu \left (
     s_j^\dag s_j + t_{\alpha j } ^\dag t_{\alpha j } - 1 
    \right ).
\end{align}
Additionally, we shall model the effect of an applied magnetic field on our triplon bands. For an arbitrary field $\textbf h = (h_x, h_y, h_z)$, the magnetic term in $H$ is 
$H_{h} = -\sum_{j} h_{\alpha}(\sigma^\alpha_{jL} + \sigma^\alpha_{jR})$, which, when expressed in terms of our triplon operators, becomes
\begin{align}
    H_h = \sum_{j}h_\alpha 2i\varepsilon_{\alpha \beta \gamma} t^\dag_\beta t_\gamma.
\end{align}

Since we expect the system to form a singlet VBS, let us introduce a real singlet condensate density order parameter $\bar s = \braket{s_j}$. The three dimers in the unit cell (shown in Fig.~\ref{fig:lattice_to_kagome}a) are equivalent by $C_3$ rotation, so we may describe all singlets with the same $\bar s$. Finally, the total Hamiltonian can be rewritten as
\begin{align}
    H &=H_0 + H_2 + H_4,
\end{align}
with
\begin{align}
    H_0 &=  N\left [ \bar s^2 (K + 2 J - \mu)  + \mu\right ],
\end{align}
\begin{align}
    \begin{split}
    H_2 &=  \sum_{j \in  \mathcal D} 
    (K - 2 J - \mu) t_{\alpha j}^\dag t_{\alpha j} 
    - (K+\mu) t_{\bar \alpha j }^\dag t_{\bar \alpha j }
    \\ 
    &\qquad \quad +h_\alpha 2i\varepsilon_{\alpha \beta \gamma} t^\dag_\beta t_\gamma \\
    &\quad - \sum_{jk \in \mathcal D, \alpha} 
    J^\alpha _{jk} \bar s^2 \left (
     t_{\alpha k}^\dag t_{\alpha j} 
    - t_{\alpha j} t_{\alpha k} + h.c.
    \right ),
    \end{split}
\end{align}
\begin{align}
    H_4 &=  - \sum_{jk \in \mathcal D, \alpha} J^\alpha _{jk} 
    \left (
    t_{bj}^\dag t_{bk} t_{ck}^\dag t_{cj} 
    - t_{bj}^\dag t_{bk}^\dag t_{cj} t_{ck} + h.c.   
    \right ),
\end{align}
where the parameter $J^\alpha_{jk}$ represents either $K$ or $J$, depending on whether $\alpha$ matches the bond colouring or not. In $H_4$, the indices $b$ and $c$ describe the two directions that do not match $\alpha$. Note that we may discard terms with three $t_\alpha$ due to reflection symmetry~\footnote{The Kagome lattice admits a reflection that swaps L and R triangles, which reverses the sign of the triple $t$ terms. Since the original spin system is invariant to this transformation we expect that these terms should vanish.}.

\begin{figure*}[t]
    \centering
    \includegraphics[width=\textwidth]{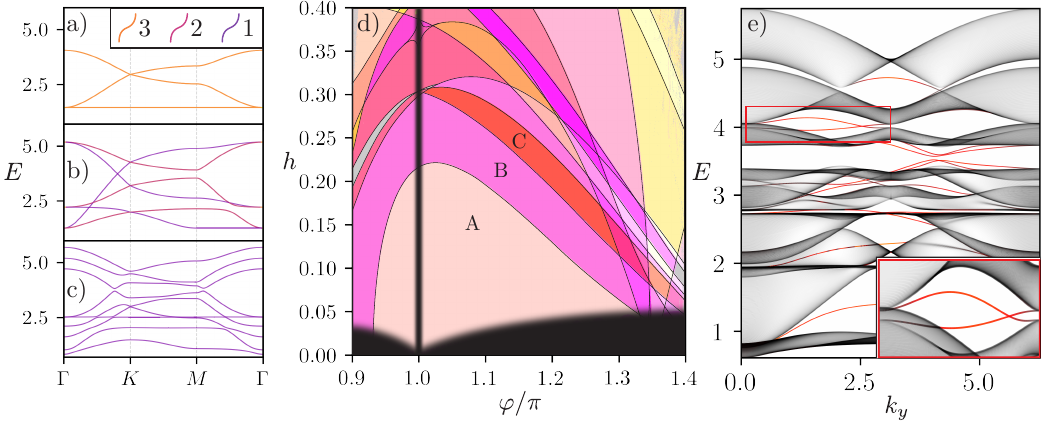}
    \caption {\textbf{a-c)} Triplon excitation band structure, with the degeneracy of each energy indicated by line colour --  a legend is provided at the top of panel a. 
    \textbf{a)} Band structure for the pure Heisenberg case ($\varphi = \pi$) with no applied field. The triple degenerate energy levels correspond to states $t_\alpha$ with $\alpha = x,y,z$. 
    \textbf{b)} The band structure for a system with some Kitaev anisotropy ($\alpha = 1.25\pi \implies K \sim -2.1, J\sim -0.71 $) and zero magnetic field. Here the triple degenerate bands are split into one double degenerate band and one non-degenerate band, however all bands remain gapless. 
    \textbf{c)} The band structure in the above case ($\varphi = 1.25\pi$), with an applied magnetic field ($h = 0.15$). Here we see that all bands are non-degenerate and gapped. 
    \textbf{d)}A plot showing the rich phase diagram of possible Chern numbers for the nine triplon bands as a function of the two parameters: Kitaev anisotropy ($\alpha$) and applied magnetic field ($h$). Regions where the bands are gapless are indicated in black and regions where the numerical process failed to converge to a stable solution are shown in grey. Each area with a distinct colouring represents a different set of the Chern numbers of the nine bands. For example the region labelled $A$ has Chern numbers $(-1, 3, -5, 6, -6, 6, -5, 3, -1)$, the $B$ region has $(-1, 3, -5, 6, -6, 3, -2, 3, -1)$, and the $C$ region has $(-1, 3, 1, 0, -6, 3, -2, 3, -1)$.
    \textbf{e)} Band structure for a strip system with periodic boundaries in the $y$-direction and open boundaries in the $x$-direction. We have set $\varphi = 1.25\pi$, $h = 0.15$ - thus are in region $B$. Bands are plotted as a function of $k_y$ for a strip containing 130 unit cells in the $x$-direction. States are coloured by their inverse participation ratio, $\textup{IPR} = \sum_{x} |\psi(x)|^4$, where red indicates a localised edge mode, and grey indicates a delocalised bulk mode. Inset shows an enlarged section of the band structure, where a pair of edge modes can be seen connecting two otherwise gapped bands. 
    }
    \label{fig:triplon_results}
\end{figure*}

Finally, the quartic terms can be decoupled by introducing the following standard mean-fields
\begin{align} \label{eqn:pqself_consistency}
    P_{jk}^\alpha &= \braket{t_{j\alpha}^\dag t_{k\alpha}} 
    \\ \label{eqn:pqself_consistency2}
    Q _{jk}^\alpha &= \braket{t_{j\alpha} t_{k\alpha}}
\end{align}
Thus, we are left with a quadratic Hamiltonian $H_{MF}(\mu, \bar s, P^\alpha_{jk} , Q^\alpha_{jk} )$ for the triplons  and the four parameters are determined self-consistently, this is detailed in appendix \ref{apx:triplon_mean_field_theory}. In practice, we found that the effect of the quartic terms, and thus the $P$ and $Q$ mean fields, was negligible. This is because the density of triplons was low across the whole phase diagram, with $\braket{t^\dag t} \approx 0.05$ on any given site, and the effect of triplon-triplon interactions was extremely weak, generally producing a $\sim 5\%$ correction to the energy of each band. Thus, in the following we shall neglect the quartic terms, and all features discussed are also present when quartic terms are included.

We now investigate the triplon excitation spectrum obtained from the self-consistently determined Hamiltonian. Since the unit cell contains 3 sites, each hosting three triplon states, the excitation spectrum will contain 9 energy levels. Provided these are gapped, the Chern number of each band can be calculated. A detailed description of the Bogoliubov-de-Gennes method and how the topological invariant are obtained is provided in Appendix~\ref{apx:triplon_mean_field_theory}.

Triplon spectra were determined using two parameters to tune the Hamiltonian. The Kitaev anisotropy (i.e.~the relative strength of $K/J$) was controlled by sweeping over a range of $\varphi \in [0.9\pi, 1.4\pi]$, approximately the range over which the system forms a stable VBS. Additionally, we applied a magnetic field in the $(1,1,1)$ direction, sweeping over a range of field strengths $h_{\alpha} = h \in [0,0.35]$. The results for these calculations are shown in Fig.~\ref{fig:triplon_results}. 

In the pure Heisenberg limit, we have three exactly degenerate triplon bands, each corresponding to a $t_x$, $t_y$ or $t_z$-type excitation, with a Dirac-like band touching at the $\Gamma$ point. In order to allow for topological excitations to form, two ingredients are needed: gaps need to be opened between the bands, and TRS must be broken to allow for non-zero Chern number. Applying a magnetic field to the Heisenberg system is not sufficient for topological bands, as a field simply splits the three degenerate triplons with an energy shift, producing a $\ket {\uparrow \uparrow}$ state aligned with the field with slightly lower energy, a $\ket {\downarrow \downarrow}$ state with slightly higher energy, and a $\ket{\uparrow \downarrow}+\ket{\downarrow \uparrow}$ band that is unaffected. Similarly, introducing only a Kitaev anisotropy $\varphi \neq \pi$ is also insufficient as the bosonic Hamiltonian retains TRS. Spectra for these cases are shown in Fig.~\ref{fig:triplon_results}a-b.

However, in the presence of both a magnetic field and the Kitaev anisotropy, the system generally has nine distinct gapped triplon bands, shown in Fig.~\ref{fig:triplon_results}c. Most interestingly, the bands have a rich phase diagram of possible Chern numbers depending on these two parameters. A full phase diagram for these is shown in Fig.~\ref{fig:triplon_results}d. With open boundaries, the bands then have chiral triplon edge modes by the bulk-edge correspondence, an example is shown in Fig.~\ref{fig:triplon_results}e.

\section{Discussion and Outlook}
\label{Discussion:Outlook}

We have studied the Kitaev-Heisenberg model on the star lattice and determined the ground state phase diagram. By tuning the relative strength of the Kitaev and Heisenberg couplings, we unearth a rich phase diagram with six distinct phases. Two of these -- the FM and stripy order -- have analogues on the corresponding honeycomb  model. Extended chiral QSL phases are found around the exact Kitaev limits, with the FM QSL being more stable than the AFM one. In both phases, we find non-zero Chern numbers of the effective Majorana bands which originate from the spontaneous breaking of TRS due to fluxes on the triangular plaquettes  of the star lattice. We confirmed the existence of chiral Majorana edge modes weakly renormalized by the presence of the Heisenberg interaction.

In the AFM range of the phase diagram we find two phases with no analogues on the honeycomb lattice, i.e. a singlet VBS and an exotic triplet VBS. These appear because the simple Neel AFM ordering that appears on the Honeycomb is geometrically frustrated on the star lattice. We unveil a symmetry between the two VBS states by generalizing the sublattice-dependent spin rotation which relates the Hamiltonian of different parts of the phase diagram. 
We then develop a bond operator formalism to show that the VBS phases have  a rich topological structure for the triplon excitations. The  introduction of a magnetic field in conjunction with the Kitaev interaction leads to gapped bands with a whole variety of (large) Chern numbers. 
Accordingly we find that the Kitaev-Heisenberg model on the star lattice cannot only support chiral Majorana excitations but  also chiral triplon edge modes.

Our work raises a number of questions for further study. First, the star lattice provides one of the simplest examples of a non-bipartite lattice. For the pure Kitaev model the presence of odd-number triangle plaquettes led to the first exactly soluble \textit{chiral} QSL with spontaneously broken  TRS~\cite{Yao_2007}, and for the pure Heisenberg limit was shown to realize a VBS ground state~\cite{Richter_2004}. Thus, an interesting avenue will be to explore whether the Kitaev-Heisenberg model shows similarly rich phase diagrams on other non-bipartite decorated lattices, possibly with exotic VBS phases. Second, the new triplet VBS phase we discover around $\varphi \approx 1.7\pi$ shows intriguing topological triplet and singlet excitations but many open questions remain. For example, in the absence of an external magnetic field the dynamics are expected to be identical to the singlet VBS phase due to the four-unit-cell rotation discussed above. However an applied magnetic field breaks this symmetry, and the possibility remains that this phase could have qualitatively very different properties than the singlet VBS. In addition, it will be important to understand the stability of the topological triplons and their experimental signatures.
Third, we showed that a sublattice dependent spin rotation can connect different parameter regimes of the Kitaev-Heisenberg model similar to the honeycomb lattice. Here on the star lattice it also uncovers a connection between two distinct VBS phases. It will be interesting to explore other lattice generalizations of such isometries, which would possibly allow us to find exact magnetic ground states even on amorphous lattices with net zero magnetization. 

Finally, there is a rich variety of modifications that could be made to the Hamiltonian in order to further assess the robustness of these different phases, especially focussing on their chiral excitations. This is particularly timely in light of recent experimental work which demonstrated the successful synthesis of a spin $1/2$ star lattice structure with antiferromagnetic couplings~\cite{sorolla_synthesis_2020}. The material is expected to have in-triangle couplings much stronger than the inter-triangle couplings. Thus it would be worth considering the phase diagram when the relative strength of couplings is adjusted. In addition, it would be interesting to derive microscopic Hamiltonians based on the microscopic orbital configuration and ab-initio calculations, e.g.~considering the effect of including symmetric off-diagonal exchanges~\cite{Rau_2014, Knolle_2018}, as well as Dzyaloshinskii-Moriya interactions \cite{kim_realization_2016, zhang_interplay_2021}.

In conclusion, the two limits of the pure Kitaev and Heisenberg models on the star lattice have been known to realize sought-after quantum magnentic phases, e.g.~a chiral QSL and a singlet VBS. We have shown that the phase diagram of the full Kitaev-Heisenberg model is much richer and allows to tune from a phase with chiral Majorana excitations to an exotic bond-selective triplet VBS phases with chiral triplon excitations. We expect that  competing exchange interactions on other decorated lattices can lead to similarly rich physics and hope that material synthesis will help with their realisation. 

\vspace{1.cm}

{\it Acknowledgements---} We acknowledge support from the Deutsche Forschungsgemeinschaft (DFG, German Research Foundation) under Germany’s Excellence Strategy–EXC–2111–390814868 and TRR 360 – 492547816. JK acknowledges support via the Imperial-TUM flagship partnership. The research is part of the Munich Quantum Valley, which is supported by the Bavarian state government with funds from the Hightech Agenda Bayern Plus. This work was supported in part by the Engineering and Physical Sciences Research Council, grant number: EP/R513052/1. This research was supported in part by the International Centre for Theoretical Sciences (ICTS) for the program "Frustrated Metals and Insulators" (code: ICTS/frumi2022/9).

\bibliography{refs}
\appendix

\section{iDMRG Results} \label{apx:DMRG_results}

\begin{figure*}[t]
    \centering
    \includegraphics[width = \textwidth]{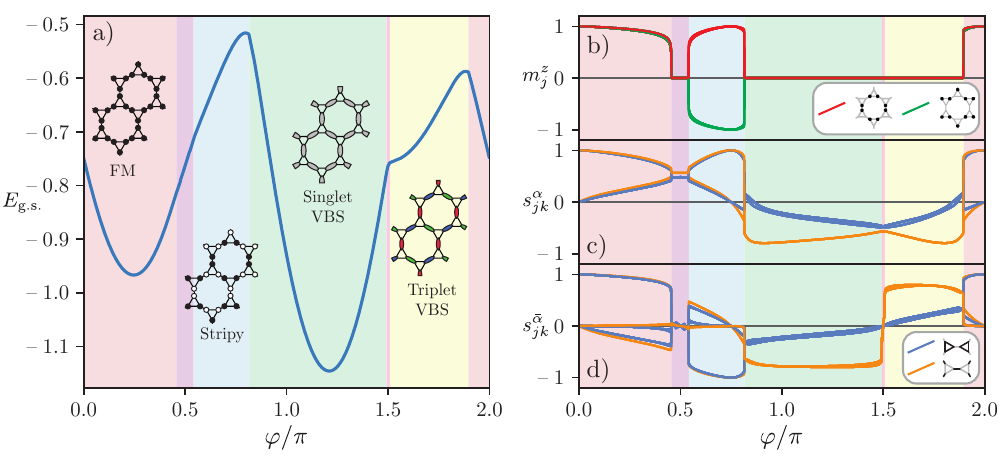}
    \caption { 
    The iDMRG results for the star lattice KH model as a function of the parameter $\varphi$, defined in Eqn.~\ref{eqn:phi_def}. Phase boundaries are indicated with background colour.
    \textbf{a)} Ground state energy per site as a function of $\varphi$.  The four non-KSL phases are shown with a diagram of the corresponding phase. KSL phases appear at $\varphi = 0.5\pi, 1.5\pi$.
    \textbf{b)} Expectation value of the $z$-component of magnetisation for each site on the lattice $m_z = \braket{\sigma_j^z}$.
    \textbf{c)} Spin-spin correlations for each bond in the \textit{on-colouring} direction, $\braket{\sigma_j^\alpha \sigma_k^\alpha}$.
    \textbf{d)} Spin-spin correlations for each bond in the two remaining \textit{off-colouring} directions, $\braket{\sigma_j^\alpha \sigma_k^\alpha}$.  In (c) and (d), correlations for in-triangle bonds are coloured in blue, whereas inter-triangle bonds are shown in orange.  
    }
    \label{fig:dmrg_results}
\end{figure*}

Here we present the results of the iDMRG computation used to numerically determine the ground-state phase diagram. The calculation was performed using the TeNPy library \cite{tenpy} using a $2\times2$ unit cell fragment of the star lattice, with 12 sites wrapping the system in the finite direction and 24 sites in total. The ground state, represented with a matrix product state ansatz with bond dimension $\chi = 1600$, was calculated for 1000 values of $\varphi \in [0,2\pi]$, starting from an initial product state. Two initial states were used, a ferromagnetically aligned state and an anti-aligned state corresponding to the stripy AFM ground state. In each case we determine the expectation value of the magnetisation in the $z$-direction,
\begin{align}
    m^z_j = \braket{\sigma_j^z}
\end{align}
as well as the expectation value of the $x$, $y$ and $z$-spin-spin correlations for each bond,
\begin{align}
    s_{jk}^\alpha = \braket{
    \sigma_j^\alpha 
    \sigma_k^\alpha }.
\end{align}
The results are shown in Fig.~\ref{fig:dmrg_results}. The phases that comprise the ground state can be identified by considering these observables. 

In the ferromagnetic phase all spins are aligned in the $+z$ direction. In the stripy phase, half of the spins have have $m^z = +1$ and the other half have $m_z = -1$. The values of $s_{jk}^\alpha$ can then be inferred directly from the $m^\alpha$ magnetisations, indicating that the system is in a magnetically ordered phase.

In the VBS and QSL phases, the system has zero net magnetisation on all axes, so all $m^\alpha$ vanish. In both the VBS phases, the spin correlations on inter-triangle bonds are much stronger than that on the in-triangle bonds, suggesting that the VBS is dimerising on the inter-triangle bonds. In the singlet VBS (around $\varphi = \pi$), both the on and off-colouring correlations are negative on inter-triangle bonds, indicating that the spins are anti-aligned in all three axes, forming a singlet. In the triplet VBS spins anti-align in the on-colouring direction and align in the two off-colouring directions, which is consistent with the triplet VBS phase described in \textsection\ref{sec:dmrg}.

\section{Majorana Mean Field Study}\label{sec:mft}
\begin{figure*}[t]
    \centering
    \includegraphics[width = \textwidth]{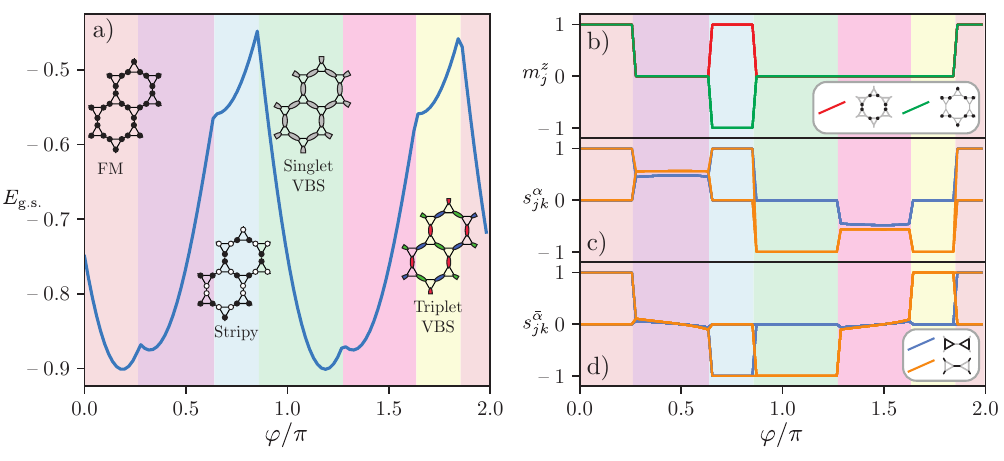}
    \caption { 
    The Majorana MFT results for the star lattice KH model as a function of the parameter $\varphi$, defined in Eqn.~\ref{eqn:phi_def}. Phase boundaries are indicated with background colour.
    \textbf{a)} Ground state energy per site as a function of $\varphi$.  The four non-KSL phases are shown with a diagram of the corresponding phase. KSL phases appear at $\varphi = 0.5\pi, 1.5\pi$.
    \textbf{b)} Expectation values of the  $z$-component of magnetisation for each site on the lattice $m_z = \braket{\sigma_j^z}$, found using the magnetically ordered mean field channel. Line colour indicates the site position with respect to arrangement of spins in the stripy phase.
    \textbf{c)} Spin-spin correlations for each bond in the \textit{on-colouring} direction, $s_{jk}^\alpha = \braket{\sigma_j^\alpha \sigma_k^\alpha}$, where we sum over the correlations calculated with both mean field channels..
    \textbf{d)} Spin-spin correlations for each bond in the two remaining \textit{off-colouring} directions, $\braket{\sigma_j^\alpha \sigma_k^\alpha}$, calculated as in part c.  In (c) and (d), correlations for in-triangle bonds are coloured in blue, whereas inter-triangle bonds are shown in orange.    
    }
    \label{fig:mft_results}
\end{figure*}

In this appendix we introduce a Majorana parton mean field theory (MFT) description of the physics involved. The uses of this method are twofold, first we shall corroborate the ground state phase diagram determined in \textsection\ref{sec:dmrg} -- providing evidence that the results are robust for large system sizes. Secondly, the MFT will allow us to characterise the topological nature of excitations in the the KSL phases.

We construct the parton description following Kitaev~\cite{Kitaev_2006}. Each spin is represented with four Majorana operators $(c\, b^x\, b^y\, b^z)$, such that $\sigma_j^\alpha \rightarrow i c_j b_j^\alpha$. Under this transformation, the Hamiltonian takes the form
\begin{align} \label{eqn:maj_hamiltonian}
    H = - \sum_{\langle jk \rangle_\alpha}
    \bigg [  
    K (i b^\alpha_j   c_j)  i b^\alpha_k c_k  
    + J \sum_{\bar \alpha \neq \alpha} 
    (i b^{\bar \alpha}_j   c_j )  ib^{\bar \alpha}_k c_k 
    \bigg ].
\end{align}
In order to apply a mean field treatment, the Majorana operators must be paired up into fermionic bilinears, which can then be replaced with their ground state expectation value -- resulting in a quadratic Majorana Hamiltonian. Of course, there are several different ways that the operators can be paired up, each giving a different MFT channel. We shall consider two physically-motivated channels for decoupling the Hamiltonian, a spin liquid channel and a magnetically ordered channel.

{\it The Spin Liquid Channel:---} In the Kitaev limit ($\varphi = 0.5\pi $ or $1.5\pi$) $J$ vanishes completely, leaving only the `on-colouring' $K$-part of the Hamiltonian. Here, the Hamiltonian has an extensive set of symmetries, since each $b_j^\alpha$ operator appears only once in $H$. Thus, there is an extensive set of fermionic operators $\hat u_{jk}^\alpha = ib_j^\alpha b_k^\alpha$, that commute with each other and the Hamiltonian. These determine a set of static $\mathbb Z_2$ gauge fields, allowing for an exact solution to be constructed~\cite{Kitaev_2006}.

Once we stray from the exact Kitaev limit, off-colouring bond operators are introduced that break the exactly commuting structure of the $\hat u_{jk}^\alpha$ operators, meaning that the $\mathbb Z_2$ fluxes are no longer conserved and no exact solution exists. However, at least in the limit of $J \ll K$ we expect that the spin liquid phase will persist, and propose a mean field decoupling that can capture this state~\cite{Schaffer_2012,Knolle_2018}. We start by rearranging the Hamiltonian to the form
\begin{align}
    H = \sum_{{\langle jk \rangle}_{\alpha}}
    \bigg [  
    K (i b^\alpha_j b^\alpha_k) i c_j c_k  
    + J \sum_{\bar \alpha \neq \alpha} 
    (i b^{\bar \alpha}_j b^{\bar \alpha}_k) i c_j c_k  
    \bigg ],
\end{align}
and introducing the following two sets of mean field parameters.
\begin{align}
    \langle i c_j c_k \rangle = \chi_{jk} \label{eqn:c_mean_field}\\ 
    \langle i b^\alpha_j b^\alpha_k \rangle = u_{jk}^\alpha.\label{eqn:b_mean_field}
\end{align}
Using the standard approximation $\hat U \hat V \approx \langle U \rangle \hat V + \hat  U \langle V \rangle  -  \langle U \rangle  \langle V \rangle$, we transform the Hamiltonian to the form
\begin{align}\label{eqn:mean_field_spin_liquid_ham}
    \begin{aligned}
    H_{SL} = \sum_{{\langle jk \rangle}_{\alpha}}
    \big[
     &\left(
    K u^\alpha_{jk} +J u^{\bar \alpha}_{jk}
    \right) 
    i c_j c_k \\
    &+ \chi_{jk} 
    ( Ki b^\alpha_j b^\alpha_k + J  ib^{\bar \alpha}_j b^{\bar \alpha}_k)\big] \\
    - \sum_{{\langle jk \rangle}_{\alpha}}
    &( K u^\alpha_{jk} + J u^{\bar \alpha}_{jk}  ) \chi_{jk}
    .    
    \end{aligned}
\end{align}
Note that here and in the following discussion, we will assume summation over the two off-colouring  spin operators labelled by $\bar \alpha$.

{\it The Magnetically Ordered Channel:---} In the ferromagnetic case ($\varphi = 0$), we expect the ground state to magnetically align, with some non-zero expectation value for $\braket{\sigma^\alpha_j} = \braket{i c_j b^\alpha_j}$. Clearly this is not captured by the above mean field theory, so we introduce a second channel using the parameters
\begin{align} \label{eqn:m_mean_field}
    \langle i b^\alpha_j   c_j \rangle = m^\alpha _j.
\end{align}
Under this parametrisation, the Hamiltonian is decomposed into the following form,
\begin{align} \begin{aligned}
    H_{M} = - \sum_{\braket{jk}} &\big [  
    K \left ( 
    im^\alpha_j b^\alpha_k c_k  + 
    ib^\alpha_j c_j m^\alpha_k  
    \right )   \\ 
    &  + 
    J \left ( 
    im^{\bar \alpha}_j b^{\bar \alpha}_k c_k  + 
    ib^{\bar \alpha}_j c_j m^{\bar \alpha}_k  
     \right ) 
    \big] \\ 
    + \sum_{\braket{jk}} 
    & \left [ 
    K m^\alpha_j m^\alpha_k 
    + J m^{\bar \alpha}_j m^{\bar \alpha}_k 
    \right ].
\end{aligned}\end{align}
Overall, our final mean field Hamiltonian is given by the sum of these two decouplings as $H_{MF} = H_{SL} + H_{M}$.

The resulting quadratic Hamiltonian is parametrised by the three sets of mean fields, and can be written in terms of the Majorana vector $\Psi = (\bm c, \bm b^x, \bm b^y, \bm b^z)^T$. Here $\bm c = (c_0, c_1 ... c_N)$, with an equivalent description for $\bm b^\alpha$. Thus the overall Hamiltonian may be expressed as
\begin{align}
    H_{MF}(\chi, u, m) = \Psi^T iA(\chi, u, m)  \Psi,
\end{align}
where $A$ is a real skew-symmetric matrix determined by the mean fields. The ground state expectation value of the mean fields can be determined by diagonalising the matrix $iA$, finding the matrix projector onto the negative eigenstates $P$, and then using the identity
\begin{align} 
    \braket{i \hat \psi_j \hat \psi_k} = -2i P_{jk},
\end{align}
where $\hat \psi_j$ is the corresponding Majorana operator at position $j$ in the Majorana vector $\Psi$.

Thus, a self-consistent MFT can be determined for a given Hamiltonian by starting with an ansatz for the three mean fields. The quadratic Hamiltonian, $H_{MF}(\chi, u, m)$, is then created and the ground state is calculated. Next, updated values for the mean fields are calculated according to eqns.~\ref{eqn:c_mean_field}, \ref{eqn:b_mean_field} and \ref{eqn:m_mean_field}. Given a new set of mean fields, one can define a new Hamiltonian, and the process can be repeated. This is iterated until a self-consistent set of mean fields is found, where their values no longer change under such an update. This was performed for 100 $\varphi$ values across the range $[0, 2\pi]$, each time starting from a set of initial ans\"atze and selecting the final converged state with the lowest energy. In each case, we tested two magnetically ordered ground state ans\"atze -- ferromagnetic and stripy -- and two spin liquid ans\"atze -- ferromagnetic and antiferromagnetic. However in general we note that results often depended on the choice of initial guess -- starting with an unphysical initial state, one often converges to an unphysical final state -- some physical intuition must be used to ensure the right states are chosen.

Results are shown in Fig.~\ref{fig:mft_results}, where we plot the ground state energy, the on-site $z$-magnetisation, $m^z_j$, and the spin-spin correlations, which are summed over both mean field channels,
\begin{align}
     s_{jk}^\alpha \coloneqq \braket{\sigma^\alpha_j \sigma_k^\alpha} = m^\alpha_j m^\alpha_k - \chi_{jk}u^\alpha_{jk}.
\end{align}
Note that the system always converged to a ground state where one of the two decoupling channels vanishes.

Six phases are found, with the phase boundaries shown in Fig.~\ref{fig:mft_results}, as well as Fig.~\ref{fig:lattice_and_phase_diagram}b. Four of these phases -- the ferromagnetic and stripe order, as well as the two chiral spin liquid phases -- have analogues in the honeycomb KH model. For $\varphi \in [0.83\pi, 1.26\pi]$ a singlet valence bond solid appears. Here, the mean fields on all triangle bonds vanish completely and on inter-triangle bonds we have $\braket{\sigma^\alpha_j \sigma_k^\alpha} = -\chi_{jk} u^\alpha_{jk} = -1$ for all $\alpha$, with $m = 0$, indicating that the spins are fully anti-aligned into a singlet. For  $\varphi \in [1.63\pi, 1.83\pi]$, a similar VBS appears with non-zero fields only on the inter-triangle bonds. However in this case we have $\braket{\sigma^\alpha_j \sigma_k^\alpha} = -1$ for $\alpha$ matching the bond colouring, and $\braket{\sigma^{\bar \alpha}_j \sigma_k^{\bar \alpha}} = +1$ for the two off-colouring directions, ${\bar \alpha}$. This is consistent with a triplet valence bond solid, where on each bond a triplet state forms whose direction depends on the bond colouring. 

Now we turn our attention to the topological nature of the Kitaev phases around $\varphi = \pi/2$ and $ 3\pi/2$. Here, the $m$-fields vanish, as well as the off-colouring $u$-fields. This means that the mean field Hamiltonian is block-diagonal, with the $c_j$ Majoranas and three sets of $b_j^\alpha$ Majoranas completely decoupled from one another. Thus, we generically have 12 sets of energy levels. The ground state Chern number is encoded in the $c$-type Majorana spectrum. 

In both Kitaev phases, the system has two degenerate ground states related by a sign flip of all $u$ fields. This is a generic feature of the Kitaev model on any lattice with odd cycles~\cite{Yao_2007,Cassella_2023}. The Chern number of the ground state can be calculated from the projector $P$ onto the negative energy eigenstates of the matrix $iA$. Here, the two degenerate states have Chern numbers $\pm 1$. In open boundaries, this means that chiral edge modes form at the boundaries. An example is shown in fig.~\ref{fig:kitaev_chern_bands}.

\section{Triplon Mean Field Theory}\label{apx:triplon_mean_field_theory}

In this section we shall provide a detailed explanation of the methods used for solving the triplon Hamiltonian using bosonic triplon MFT. It has the advantage that we can directly obtain the topology of the triplon bands. This section has three parts: First we discuss the momentum space construction of the Hamiltonian, next we shall explain the methods used to solve this Hamiltonian for the bosonic eigenstates. Finally we derive the form of the Chern number for a bosonic system. 

\subsection{Construction of the Hamiltonian}

Let us start by restating the Hamiltonian derived in \textsection\ref{sec:bond_operator}, given by 
\begin{align}
    H &=H_0 + H_2 + H_4,
\end{align}
where
\begin{align}
    H_0 &=  N\left [ \bar s^2 (K + 2 J - \mu)  + \mu\right ],\\
    \begin{split}
    H_2 &=  \sum_{j \in  \mathcal D} 
    (K - 2 J - \mu) t_{\alpha j}^\dag t_{\alpha j} 
    - (K+\mu) t_{\bar \alpha j }^\dag t_{\bar \alpha j }
    \\ 
    &\qquad \quad +h_\alpha 2i\varepsilon_{\alpha \beta \gamma} t^\dag_\beta t_\gamma \\
    &\quad - \sum_{jk \in \mathcal D, \alpha} 
    J^\alpha _{jk} \bar s^2 \left (
     t_{\alpha k}^\dag t_{\alpha j} 
    - t_{\alpha j} t_{\alpha k} + h.c.
    \right ).
    \end{split}
\end{align}
The four-boson Hamiltonian is decoupled using the mean fields, 
\begin{align} 
    P_{jk}^\alpha &= \braket{t_{j\alpha}^\dag t_{k\alpha}} 
    \\ 
    Q _{jk}^\alpha &= \braket{t_{j\alpha} t_{k\alpha}},
\end{align}
taking the following form
\begin{align}\begin{aligned}
    H_4 = 
    & - \sum_{jk \in \mathcal D, \alpha} J^\alpha _{jk} 
    \Big (
    P_{jk}^{b} t_{cj}^\dag t_{ck} +  t_{bj}^\dag t_{bk} P_{jk}^{c*} + h.c.  
    \\
    & \qquad 
    - t_{bj}^\dag t_{bk}^\dag Q^{c}_{jk} -  Q^{b * }_{jk} t_{cj} t_{ck}
    + h.c.   
    \Big )\\
    & + \sum_{jk \in \mathcal D, \alpha} J^\alpha _{jk} 
    \Big (
        P_{jk}^b P_{jk}^{c*} - Q^{b * }_{jk} Q^{c}_{jk} + h.c.
    \Big ).
\end{aligned} \end{align}
Note that here, the indices $b$ and $c$ denote the two subscripts that are not equal to $\alpha$ for each term in the sum. 
Since we are working in a translationally invariant system, we transform to momentum space. We replace every position index $j$, with a pair $\textbf r, j$, where $\textbf r$ labels the unit cell position and $j$ now labels the site within the unit cell. Momentum-space annihilation operators are defined as
\begin{align}
    t_{j \textbf r , \alpha} =\frac{1}{\sqrt N} \sum_{\textbf q} e^{-i \textbf r \cdot \textbf q }t_{j \textbf q , \alpha}.
\end{align}
Under this, terms in the Hamiltonian containing triplet operators are transformed according to
\begin{align}
    \sum_{\textbf r} 
    t^\dag_{j \textbf r + \bm \Delta , \alpha} 
    t_{k \textbf r, \alpha} & = 
    \sum_{\textbf q} e^{i \textbf q \bm \Delta} 
    t_{j \textbf q, \alpha}^\dag
    t_{k \textbf q, \alpha}, 
\end{align}
\begin{align} \begin{split}
    \sum_{\textbf r} 
    t_{j \textbf r + \bm \Delta , \alpha} 
    t_{k \textbf r, \alpha} & = 
    \frac{1}{2}
    \sum_{\textbf q} \Big (
    e^{i \textbf q \bm \Delta} 
    t_{j -\textbf q, \alpha}
    t_{k \textbf q, \alpha} \\
    & \qquad \ + 
    e^{-i \textbf q \bm \Delta} 
    t_{k -\textbf q, \alpha}
    t_{j \textbf q, \alpha} 
    \Big ).
\end{split}
\end{align}
Thus we get a full Hamiltonian containing the following seven sets of terms:
\begin{widetext}
    \begin{align}
        \textup{Constant shift: } 
        & 3\bar s^2 \left ( K + 2J -\mu \right ) + 3\mu 
        \\ 
        \textup{On-colouring on-site potential: } 
        & (K - 2J - \mu) \sum_{\textbf q, j} 
        t^\dag _{j \textbf q, \alpha }  t_{j \textbf q, \alpha } 
        \\
        \textup{Off-colouring on-site potential: } 
        & (- K - \mu) \sum_{\textbf q, j} 
        t^\dag _{j \textbf q, \alpha }  t_{j \textbf q, \alpha } 
        \\ 
        \textup{Magnetic Field: } 
        & 2ih_\alpha \sum_{\textbf q,j}\varepsilon_{\alpha \beta \gamma} t^\dag_{j\textbf q,\beta} t_{j\textbf q,\gamma} 
        \\
        \begin{split}
        \textup{Hopping: } & -
        \sum_{\braket{j,k},\alpha} 
        J^\alpha _{jk} \bar s^2 
        \sum_{\textbf q} e^{i \textbf q \cdot \bm \Delta}
        t_{j \textbf q, \alpha}^\dag t_{k \textbf q, \alpha}  + h.c.
        \\
        & +\sum_{\braket{j,k},\alpha} 
        \frac{1}{2} J^\alpha _{jk}  \bar s^2 \sum_{\textbf q,j,k} 
        e^{i \textbf q \cdot \bm \Delta}
        t_{j -\textbf q, \alpha}t_{k \textbf q, \alpha} + 
        e^{-i \textbf q \cdot \bm \Delta}
        t_{k -\textbf q, \alpha}t_{j \textbf q, \alpha} + h.c.
    \end{split}
    \end{align}
    Adding in the mean fields (i.e. the quartic interactions) we get a few additional terms:
    \begin{align}
        \textup{Constant shift: } &  
        \sum_{\textup{bonds}}\sum_{\alpha} J^\alpha _{jk} \left(
        P_{jk}^{b}P_{jk}^{c*}- 
        Q_{jk}^{b}Q_{jk}^{c*} + h.c.
        \right)
    \\
        \begin{split}
        \textup{Hopping: } &
        -\sum_{\textbf q,\alpha}
        J^\alpha _{jk} e^{i \textbf q \cdot \bm \Delta}
        \left( P^{b*}_{jk} t^\dag_{j \textbf q, c}t_{k \textbf q,c} + P^{c*}_{jk} t^\dag_{j \textbf q, b}t_{k \textbf q,b}\right) + h.c. \\
        & + \sum_{\textbf q,\alpha} \frac{1}{2} J^\alpha _{jk}
        Q_{jk}^{b*} \left ( 
        e^{i \textbf q \cdot \bm \Delta} t_{j -\textbf q, c}\, t_{k \textbf q, c}
        + e^{-i \textbf q \cdot \bm \Delta} t_{k -\textbf q, c}\, t_{j \textbf q, c}
        \right ) + h.c. \\
         & + \sum_{\textbf q,\alpha}\frac{1}{2} J^\alpha _{jk}
        Q_{jk}^{c*} \left ( 
        e^{i \textbf q \cdot \bm \Delta} t_{j -\textbf q, b}\, t_{k \textbf q, b}
        + e^{-i \textbf q \cdot \bm \Delta} t_{k -\textbf q, b}\, t_{j \textbf q, b}
        \right ) + h.c. 
    \end{split}
    \end{align}
\end{widetext}

A valid ground state is given by choosing $\mu$ and $\bar s$ such that the Hamiltonian satisfies the saddle point equations
\begin{align} \label{eqn:triplon_saddle_point}
    \left \langle
        \frac
        {\partial H_{MF}}
        {\partial \mu} 
    \right \rangle
    =
    \left \langle
        \frac
        {\partial H_{MF}}
        {\partial \bar s} 
    \right \rangle
    =0,
\end{align}
where the expectation value is calculated in the ground state. Furthermore, $P$ and $Q$ must satisfy the self consistency relations (Eqns.~\ref{eqn:pqself_consistency} and \ref{eqn:pqself_consistency2}). Self-consistent values for $\mu, \bar s, P$ and $Q$ are determined by using a root finding algorithm (\texttt{scipy.optimize.root}) to converge to a set of mean fields that satisfy these conditions. At each step of the iteration, the Hamiltonian is solved using the bosonic Bogoliubov-de-Gennes method~\cite{Ripka_1986}, and the mean fields are determined from the bosonic ground state.  

\subsection{Bosonic Bogoliubov-de Gennes Method}

In order to solve this Hamiltonian, we shall use the bosonic Bogoliubov-de Gennes (BDG) method, detailed in \cite{Ripka_1986}. This method is well-understood, however for the sake of completeness, and to introduce the notation used to derive the Chern number in \textsection\ref{apx:chern_number}, we discuss the method here.

Let us work with a generic set of momentum-space bosonic annihilation operators, $b_{\textbf q j}$, satisfying the bosonic commutation relations $[b_{\textbf q j}, b_{\textbf p k}^\dag] = \delta(\textbf q - \textbf p) \delta_{jk}$, where the Hamiltonian is given by
\begin{align} 
    \begin{split}
        H_{\textbf q} & = 
        h_{jk}(\textbf q) b_{\textbf q j}^\dag b_{\textbf q k}  \\ 
        & + \frac{1}{2}\left ( \Delta_{jk} (\textbf q) b_{\textbf q j}^\dag b_{-\textbf q k}^\dag
        + \Delta_{jk}^\dag (\textbf q) b_{-\textbf q j} b_{\textbf q k}
        \right ), 
    \end{split}
\end{align}
where $h$ is a hermitian matrix. We assume that $h$ satisfies the relation $h^*(\textbf q) = h(-\textbf q)$, with the same for $\Delta$. Using the bosonic commutation relations, we may rewrite the first term in the following form,
\begin{align}
\begin{split}
    h_{jk}(\textbf q) b_{\textbf q j}^\dag b_{\textbf q k}  =
    & \frac{1}{2} \left ( 
    h_{jk}(\textbf q) b_{\textbf q j}^\dag b_{\textbf q k} + 
    h_{jk}^T(\textbf q)  b_{\textbf q j} b_{\textbf q k}^\dag
    \right ) \\ 
    & - \frac{1}{2} \Tr h.
\end{split}
\end{align}
Thus, the Hamiltonian can be written in a generic BdG form as 
\begin{align}
    H_{\textbf q} =
    \frac{1}{2}
    \Psi_{\textbf q}^\dag
    M(\textbf q)
    \Psi_{\textbf q}
    - \frac{1}{2} \Tr h,
\end{align}
where we have introduced the bosonic vector of creation/annihilation operators
\begin{align}
    \Psi_{\textbf q} = (b_{\textbf q 0},b_{\textbf q 1},...,b^\dag_{-\textbf q 0},b^\dag_{-\textbf q 1},...  )^T
\end{align}
and defined the BdG matrix
\begin{align}\label{eqn:m_definition}
    M(\textbf q) = 
    \begin{pmatrix}
        h(\textbf q) & \Delta  ( \textbf q) \\
        \Delta^\dag(\textbf q) & h (\textbf q)
    \end{pmatrix}.
\end{align}
To diagonalise this, let us consider the time-dependence of $\Psi_{\textbf q}$, given by
\begin{align} 
    \partial_t \Psi_{\textbf q} = i\left [ H_{\textbf q}, \Psi_{\textbf q}\right ].
\end{align}
Applying the bosonic commutation relations, we arrive at the following expression,
\begin{align}\label{eqn:boson_eom}
    \partial_t \Psi_{\textbf q} = i 
    \eta M_{\textbf q}
    \Psi_{\textbf q},
\end{align}
where the `metric' $\eta$ is defined as
\begin{align}
    \eta := 
    \begin{pmatrix}
        \1 & \\
        & -\1
    \end{pmatrix}
\end{align}
We now define a rotated set of bosonic operators $\bm \alpha_{\textbf q}$, with $\Phi_{\textbf q} = (\bm \alpha_{\textbf q 0},...,\bm \alpha^\dag_{-\textbf q 0},...  )^T$, according to
\begin{align} \label{eqn:uv_define}
    b_{\textbf qj} & = 
    \textbf u_j \cdot  \bm \alpha_{\textbf q} +
    \textbf v_j \cdot \bm \alpha^\dag_{-\textbf q},  \\
     b_{-\textbf q j}^\dag & = 
    \bar {\textbf v}_j \cdot \bm \alpha_{\textbf q}  +
    \bar {\textbf u}_j \cdot \bm \alpha^\dag_{-\textbf q}, \\
    \implies      
    \Psi_q
    & = R_{\textbf q} \Phi_{\textbf q} \textup{, for } R_{\textbf q} = \begin{pmatrix}
        U & V\\
         \bar {V} &\bar {U}
    \end{pmatrix}.
\end{align}
Here $R_{\textbf q}$ is a rotation matrix that must be determined. In order to preserve bosonic commutation relations for $\bm \alpha$, we require that $R_{\textbf q}$ satisfies
\begin{align}
    R\eta R^\dag\label{eqn:gamma_r0}
    = \eta, \\
    R^\dag \eta R\label{eqn:gamma_r}
    = \eta.
\end{align}
After applying this transformation to our equation of motion \ref{eqn:boson_eom}, we arrive at
\begin{align}
    i\partial_t \Phi_{\textbf q} = 
    R ^{-1}
    \eta M(\textbf q)
    R
    \Phi_{\textbf q}.
\end{align}
Thus, eigenstates of the Hamiltonian are found by choosing a transformation $R$ that diagonalises the operator $\eta M(\textbf q)$, which we write in the form
\begin{align}
    R^{-1} \eta M_{\textbf q} R = \eta D
\end{align}
where $D$ should have only positive eigenvalues, ensuring that the bosonic modes all have positive energy. In practice, this can be done by diagonalising the matrix $\eta M$ and adjusting the normalisation of the eigenstates such that they satisfy eqns.~\ref{eqn:gamma_r0} and \ref{eqn:gamma_r}.

Now, let us rewrite the original Hamiltonian as
\begin{align}
    H_{\textbf q} = \frac{1}{2}
    \begin{pmatrix}
        \bm \alpha^\dag_{\textbf q} &
        \bm \alpha_{-\textbf q}
    \end{pmatrix}
    R^\dag M_{\textbf q} R
    \begin{pmatrix}
        \bm \alpha_{\textbf q} \\
        \bm \alpha^\dag_{-\textbf q}
    \end{pmatrix}- \frac{1}{2} \Tr h.
\end{align}
By rearranging eqn.~\ref{eqn:gamma_r} to get the identity $R^\dag = \eta R^{-1} \eta$, and inserting it into the above, we can show that $R^\dag M_{\textbf q} R = D$ and the Hamiltonian takes the form
\begin{align}
    H_{\textbf q} = \frac{1}{2}
    \sum_{j} \left ( 
    \omega_{j} \alpha_{\textbf q j}^\dag \alpha_{\textbf q j}
    + \omega'_{j} \alpha_{- \textbf q j}\alpha_{ - \textbf q j}^\dag 
    \right ) - \frac{1}{2} \Tr h.
\end{align}
where we have labelled the positive eigenvalues of $\eta M$ as $\omega_j$ and the negative eigenvalues of $\eta M$ determine $\omega'_j$ (which are positive -- negative energies would indicate an instability, where the system can infinitely populate the negative energy states). We rearrange the second set of bosonic operators to get
\begin{align}\begin{split}
    H_{\textbf q} = &\frac{1}{2}
    \sum_{j} \big ( 
    \omega_{j} 
    \alpha_{\textbf q j}^\dag 
    \alpha_{\textbf q j}
     + \omega'_{j}
    \alpha_{ - \textbf q j}^\dag
    \alpha_{- \textbf q j}
    \big )\\
    & - \frac{1}{2}\Tr h + \frac{1}{2}\sum \omega ' 
\end{split}
\end{align}

Of course, in order to implement the full mean field theory, one needs not only the bosonic spectrum but also the ground state expectation values for the mean fields, $\braket{0 |b^\dag_{\textbf q j} b_{\textbf q' k}  | 0} $ and $\braket{0 |b_{\textbf q j} b_{\textbf q' k}  | 0} $. These can be extracted from the rotation matrix $R_{\textbf q}$ by transforming the expectation value to our diagonalising basis $\{\bm \alpha\}$:
\begin{align}\begin{split}
    \braket{0 |b^\dag_{\textbf q j} b_{\textbf q k}  | 0} 
    = \bra{0}
    &(
    \textbf u^*_{j} \cdot \bm \alpha_{\textbf q}^\dag + 
    \textbf v^*_{j} \cdot \bm \alpha_{-\textbf q}
    ) \\ 
    \times
    &(
    \textbf u_k \cdot  \bm \alpha_{\textbf q} +
    \textbf v_k \cdot \bm \alpha^\dag_{-\textbf q}
    ) 
    \ket{0}.
\end{split}
\end{align}
Since the ground state contains no bosons, the only term here that does not vanish is
\begin{align}
    \braket{0 |b^\dag_{\textbf q j} b_{\textbf q k}  | 0} 
    & =
    \braket{0 |
    (\textbf v^*_{j} \cdot \bm \alpha_{-\textbf q} )
    (\textbf v_k \cdot \bm \alpha^\dag_{-\textbf q})
    | 0} \\
    & = \braket{0 |
    V^*_{jm}V_{km}
    | 0} \\
    &= (V V^\dag) _{kj}.\label{eqn:P}
\end{align}
Equally, we can repeat the process for the other correlation function of interest,
\begin{align}
    \braket{0 |b_{-\textbf q j} b_{\textbf q k}  | 0} 
   = (V \bar U^\dag)_{kj}.\label{eqn:Q}
\end{align}

\subsection{Bosonic Chern Number}\label{apx:chern_number}

In order to define a bosonic version of the Chern number, let us start by restating the results of the previous section. We have a BdG matrix given by $M(\textbf q)$ -- assumed to be positive definite (defined in eqn.~\ref{eqn:m_definition}), with a set of eigenstates $\ket{\psi_j(\textbf q)}$ which satisfy
\begin{align}
    \eta M(\textbf q) \ket{\psi_j(\textbf q)} = 
    \lambda_j (\textbf q) \ket{\psi_j(\textbf q)}.
\end{align}
For a physical solution, we make the assumption that all eigenstates are real, i.e.~there is no dynamical instability. Then we can show that if $\ket{\psi_j(\textbf q)}$ is a right eigenstate of $\eta M(\textbf q)$, from the hermiticity of $M_{\textbf q}$, we must have a corresponding left eigenvector defined as 
\begin{align}
    \bra{\overline{\psi_j}(\textbf q)} = \bra{\psi_j(\textbf q)} \eta,
\end{align}
with the same eigenvalue. Now, let's assume that the eigenstates of $\eta M$ are all either positive or negative, then we can calculate the quantity
\begin{align}
    \braket{\overline{\psi_j} | \eta M |\psi_j} 
    &= \lambda_{j}
    \braket{\overline{\psi_j} | \psi_j} \\ 
    \implies
    \braket{\psi_j | M | \psi_j} 
    & = \lambda_{j}
    \braket{\overline{\psi_j} | \psi_j}.
\end{align}
The left hand side here is always positive due to the positive definiteness of $M$. This implies that positive eigenstates of $\eta M$ have positive norm, and the negative eigenstates of $\eta M$ have negative norm. Thus we see that (in agreement with condition \ref{eqn:gamma_r}) the states can be split into two sets and normalised, which we will label with $\phi_i^+$ and $\phi_i^-$ with the following norms
\begin{align}
    \braket{\phi_i^+ | \eta  |\phi_i^+} = +1 \\
    \braket{\phi_i^- | \eta  |\phi_i^-} = -1 .
\end{align} 

In this context, let us look now at finding relative phases between adjacent eigenstates in k-space. We can define the phase between two states as
\begin{align}
    \gamma = \arg \left [ \pm 
    \braket{\phi_i^\pm(\textbf k) | \eta  |\phi_i^\pm(\textbf q)}
    \right ].
\end{align}
In analogy with the standard discussion around Berry phase \cite{Asboth2016}, this quantity is not gauge invariant for a local gauge $\ket{\phi_i(\textbf q)} \rightarrow e^{i\alpha(\textbf q)}\ket{\phi_i(\textbf q)}$. Thus, let us define a gauge-invariant quantity -- the Berry phase around a plaquette in k-space,
\begin{align} 
    \begin{split}
        \gamma_B(\textbf q) = &\arg \left [
        \braket{\phi_i^\pm(\textbf q) | \eta  |\phi_i^\pm(\textbf q + \bm \delta_x)} \right .\\ 
        & \braket{\phi_i^\pm(\textbf q + \bm \delta_x) | \eta  |\phi_i^\pm(\textbf q + \bm \delta_x+ \bm \delta_y)}  \\
        & \braket{\phi_i^\pm(\textbf q + \bm \delta_x+ \bm \delta_y) | \eta  |\phi_i^\pm(\textbf q + \bm \delta_y)}\\
        & \left .\braket{\phi_i^\pm(\textbf q + \bm \delta_y) | \eta  |\phi_i^\pm(\textbf q)} \right ],
    \end{split}
\end{align}
where $\delta_x$ and $\delta_y$ represent a translation in momentum space by $2\pi /L$ in either the $x$ or $y$ direction, $L$ being the system size. Here the factor of $(\pm 1)^4$ from state normalisation clearly vanishes. Let us simplify this expression by defining the symplectic projector onto a single eigenstate,
\begin{align}
    \hat \phi_i^\pm (\textbf q) = \ket{ \phi_i^\pm (\textbf q)}\bra{ \phi_i^\pm (\textbf q)} \eta,
\end{align}
which is identical to a standard density matrix aside from the extra factor of $\eta$. Using this expression we can rewrite the Berry phase in the following form:
\begin{align} 
\begin{split}
    \gamma_{B} (\textbf q) = \arg \Tr &\left[
    \hat \phi_i^\pm (\textbf q)
    \hat \phi_i^\pm (\textbf q + \bm \delta_x) \right. \\ 
    &\left. \hat \phi_i^\pm (\textbf q+ \bm \delta_x + \bm \delta_y)
    \hat \phi_i^\pm (\textbf q + \bm \delta_y)
    \right]. 
\end{split}
\end{align}
Finally the Chern number is calculated using the usual expression
\begin{align} 
    \begin{split}
    C  = \frac{1}{2\pi}\sum_{\textbf q} \arg \Tr & \left[
    \hat \phi_i^\pm (\textbf q)
    \hat \phi_i^\pm (\textbf q + \bm \delta_x) \right. \\ 
    & \left. \hat \phi_i^\pm (\textbf q+ \bm \delta_x + \bm \delta_y)
    \hat \phi_i^\pm (\textbf q + \bm \delta_y)
    \right].
\end{split}
\end{align}

\end{document}